\documentclass[twocolumn,preprintnumbers,amsmath,amssymb]{revtex4}

\usepackage{graphicx}
\usepackage{dcolumn}
\usepackage{bm}
\usepackage{epstopdf}

\begin{document}


\title{Stable and scalable multistage terahertz-driven particle accelerator}

\author{Heng Tang$^{1,2*}$, Lingrong Zhao$^{1,2*}$, Pengfei Zhu$^{1,2}$, Xiao Zou$^{1,2}$, Jia Qi$^{3}$, Ya Cheng$^{3}$, Jiaqi Qiu$^{4}$, Xianggang Hu$^{5}$, Wei Song$^{5}$, Dao Xiang$^{1,2,6,7\dag}$ and Jie Zhang$^{1,2\dag}$}
\affiliation{%
$^1$ Key Laboratory for Laser Plasmas (Ministry of Education), School of Physics and Astronomy, Shanghai Jiao Tong University, Shanghai 200240, China \\
$^2$ Collaborative Innovation Center of IFSA (CICIFSA), Shanghai Jiao Tong University, Shanghai 200240, China \\
$^3$ State Key Laboratory of Precision Spectroscopy, East China Normal University, Shanghai 200062, China \\
$^4$ Nuctech Company Limited, Beijing, 100084, China\\
$^5$ Science and Technology on High Power Microwave Laboratory, Northwest Institute of Nuclear Technology, Xi'an, Shanxi 710024, China\\
$^6$ Tsung-Dao Lee Institute, Shanghai Jiao Tong University, Shanghai 200240, China \\
$^7$ Zhangjiang Institute for Advanced Study, Shanghai Jiao Tong University, Shanghai 200240, China \\
}
\date{\today}

\begin{abstract}
Particle accelerators that use electromagnetic fields to increase a charged particle's energy have greatly advanced the development of science and industry since invention. However, the enormous cost and size of conventional radio-frequency accelerators have limited their accessibility. Here we demonstrate a mini-accelerator powered by terahertz pulses with wavelengths 100 times shorter than radio-frequency pulses. By injecting a short relativistic electron bunch to a 30-mm-long dielectric-lined waveguide and tuning the frequency of a 20-period terahertz pulse to the phase-velocity-matched value, precise and sustained acceleration for nearly 100\% of the electrons is achieved with the beam energy spread essentially unchanged. Furthermore, by accurately controlling the phase of two terahertz pulses, the beam is stably accelerated successively in two dielectric waveguides with close to 100\% charge coupling efficiency. Our results demonstrate stable and scalable beam acceleration in a multistage mini-accelerator and pave the way for functioning terahertz-driven high-energy accelerators.  
\end{abstract}

\maketitle
Particle accelerators have played important roles in advancing science and industry, with applications ranging from high-energy colliders, synchrotron light sources, and free-electron lasers to x-ray imaging and cancer therapy machines. However, the large size and cost of conventional accelerators based on radio-frequency (rf) technology limit accessibility and are motivating development of compact accelerators \cite{LPA1, LPA2, LPA3, LPA4, LPA5, PWFA1, LPA6, LPA7, PWFA2, LPA8, PWFA3, LPA9, DLA1, DLA2, THz1, THz2, THz3, THz4, THz5, THz6, THz7}. Among the techniques, terahertz (THz)-driven acceleration has great potential \cite{THz1, THz2, THz3, THz4, THz5, THz6, THz7}. Compared to the rf accelerator, the approximately 100-fold reduction in wavelength for the THz accelerator promises significant reduction in size, but increase in breakdown threshold to the gigaelectronvolt-per-meter regime \cite{GV1, GV2}. Compared to the dielectric laser accelerator \cite{DLA1, DLA2}, the approximately 100-fold increase in wavelength facilitates the fabrication of the accelerating structure and relaxes the tolerance on electron beam transverse size, pulse width, and timing for stable acceleration. Compared to plasma acceleration \cite{LPA1, LPA2, LPA3, LPA4, LPA5, PWFA1, LPA6, LPA7, PWFA2, LPA8, PWFA3, LPA9}, the accelerating medium, i.e., dielectric-lined waveguide (DLW), is more stable and it is easier to control the beam size during staged acceleration. 
    \begin{figure*}[t]
    \includegraphics[width = 0.85\textwidth]{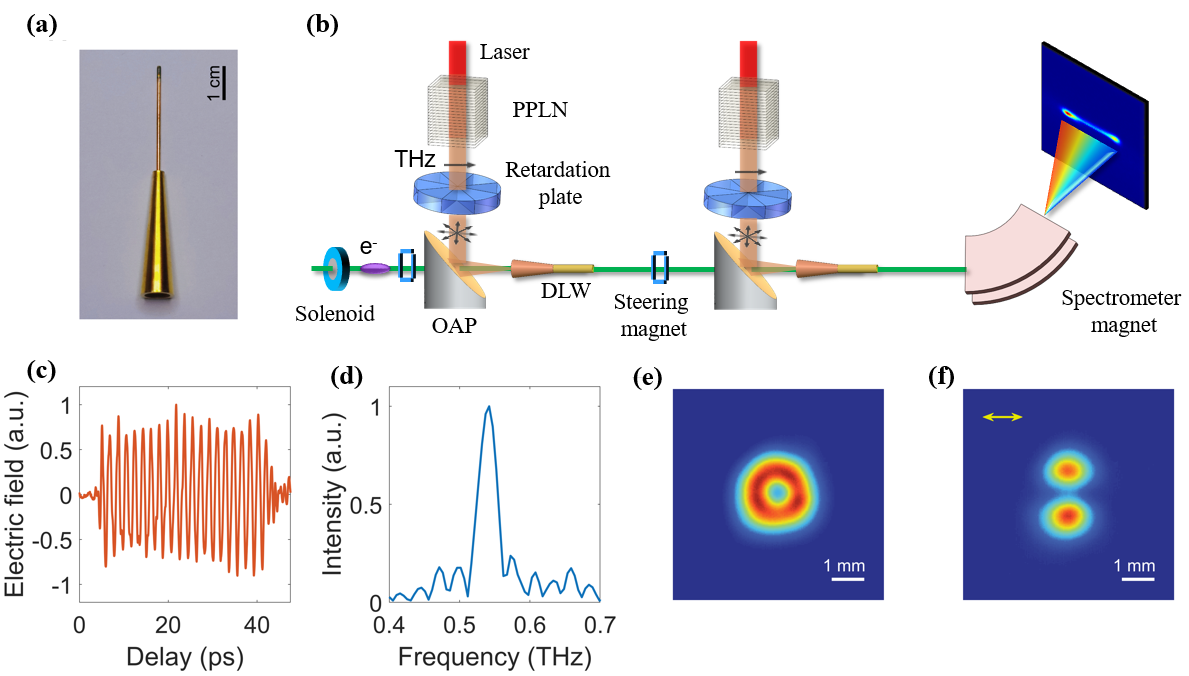}
            \caption{Mini-accelerator powered by THz pulses. (a) Picture of DLW assembled with a horn; (b) Schematic of two-stage THz acceleration experiment; Measured THz waveform (c) and spectrum (d); Measured profiles of the radially polarized THz pulse before (e) and after (f) a polarizer (the arrow representing the direction of the wire).    
    \label{Fig.1}}
    \end{figure*}

Despite extensive efforts in the past few years, challenges, such as proof-of-principle THz acceleration to a functioning THz accelerator, remain. In a functioning accelerator, the beam is stably accelerated in successive modules to attain high energy since only finite energy gain can be reached in each accelerating structure. Therefore, a functioning THz accelerator requires stable whole-bunch acceleration, i.e., all the electrons in a bunch are accelerated at approximately the same phase to reach approximately the same energy, as well as high charge coupling efficiency and accurate phase control for staging. However, with the electron beam being longer than half of the THz wavelength, only a fraction of the electrons are accelerated by a laser-generated THz pulse, while others are decelerated \cite{THz1, THz2, THz3, THz4, THz5}. Though recently, the staging of acceleration for relativistic electrons has been reported \cite{THz7}, the short acceleration length from an accelerator-produced single-cycle transition radiation THz pulse and the limited beam quality is not sufficient to address the question of whether an electron bunch can be stably accelerated in successive long structures with high charge coupling efficiency and high stability. We note that while whole-bunch acceleration for keV electron beam in two stages has been reported in \cite{THz6}, the small energy gain and large energy spread of the beam makes it difficult to answer the question of whether beam energy spread can be preserved in staged acceleration with considerable energy gain. 

In this Letter, we demonstrate sustained and stable whole-bunch acceleration of a relativistic electron beam in two stages driven by 20-period THz pulses. The accelerating structure is a 30-mm-long DLW with 0.86 mm aperture, about 100 times smaller than a conventional rf structure. The DLW consists of a dielectric layer of fused silica with a wall thickness of 43 $\mu$m and a gold-coated external surface. Our analysis shows that such a structure supports an accelerating mode at 0.54 THz with a phase velocity of approximately $v_p=0.989c$ and a group velocity of $v_g=0.72c$  (see Supplementary Materials Fig.S1). A horn with an opening angle of 10 degrees is attached to the DLW to increase the coupling efficiency \cite{THz1, THz6, THz7}, which is measured to be about 70\% in this experiment. An optical image of the mini-accelerator is shown in Fig.~1(a). The layout of the experiment is depicted in Fig.~1(b). A 3 MeV, 10 fC relativistic electron beam with a velocity of $v_e=v_p=0.989c$, is successively accelerated in two identical modules consisting of an off-axis parabolic (OAP) mirror for focusing the THz pulse, a horn for adiabatically guiding the THz pulse, and a DLW for the THz-electron interaction. A solenoid is used to control the electron beam size and steering magnets are used to control the beam trajectory during staged acceleration. The beam energy distribution is measured with a spectrometer magnet (see Fig.S2).

    \begin{figure*}[t]
    \includegraphics[width = 0.8\textwidth]{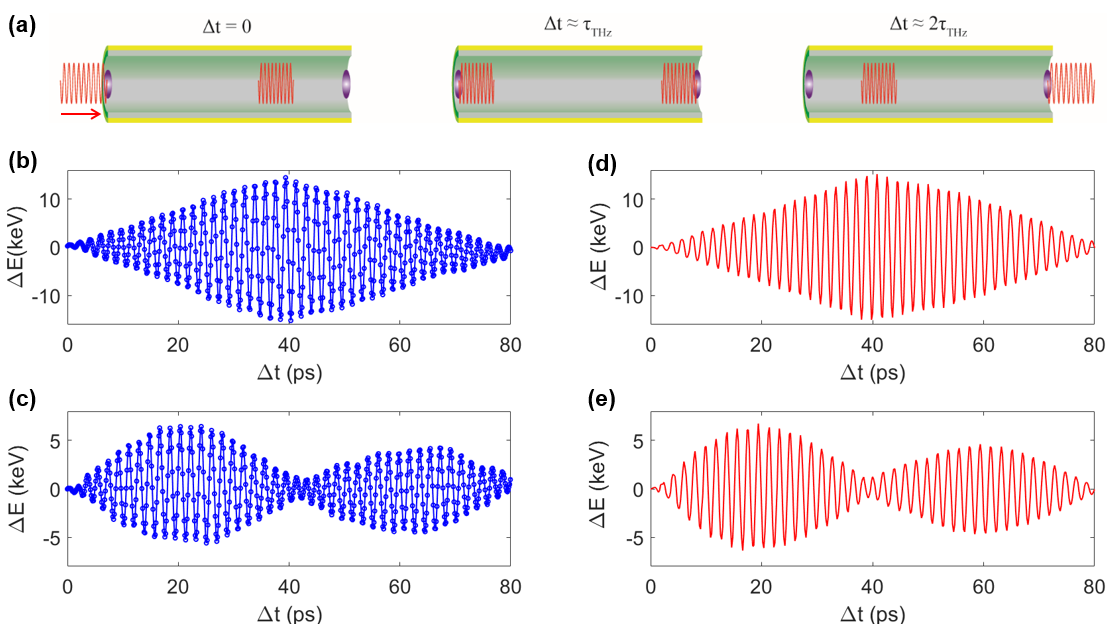}
            \caption{Visualization of THz-electron interaction. (a) Schematic depiction of the relative position between the electron beam (purple) and THz pulse (red) at various time delays (the arrow represents the direction of travel); Measured electron energy changes when the THz frequency equals to (b) and deviates from (c) that of the accelerating mode; Simulated electron energy change when the THz frequency equals to (d) and deviates from (e) that of the accelerating mode.    
    \label{Fig.1}}
    \end{figure*}

The THz pulse is produced by sending a linearly polarized femtosecond laser (about 3 mJ) through a cryogenically cooled periodically poled lithium niobate (PPLN) crystal \cite{PPLN} with a size of 3 mm by 3 mm. The PPLN crystal has a total length of 4.4 mm and a period of 0.22 mm for producing a 20-period narrowband THz pulse centered around 0.54 THz. The THz waveform measured with electro-optic sampling is shown in Fig.~1(c), and the corresponding spectrum is shown in Fig.~1(d). The duration of the THz pulse is $\tau_{THz}=40~$ps, matching the length of the DLW for maximal energy gain. The temperature of the PPLN is reduced to below 200 K for reducing THz absorption. It should be noted that the frequency of the THz pulse depends on the refractive index of the laser and THz pulse in the crystal. Therefore, the THz pulse's central frequency may be varied by changing the crystal temperature, and in this experiment this offers about 6\% tunability for the THz frequency while keeping the THz energy at a similar level (see Fig.S3). 

To allow efficient excitation of the TM01 mode in DLW, the linearly polarized narrowband THz pulse is converted into a radially polarized pulse with a spatially variable retardation plate (see Fig.S4). This conversion ensures the best matching of the Poynting vector for optimally transferring energy from the incident THz pulse to the accelerating mode in the DLW. In general, the coupling efficiency in converting the energy of a THz pulse into that of a specific mode in the DLW depends on the similarity of the Poynting vector distribution. Therefore, a linearly polarized THz pulse can efficiently excite TM11-like deflecting mode \cite{THzO} and a radially polarized THz pulse is required to excite the TM01 accelerating mode. The radially polarized THz pulse profile is measured with a THz camera and shows the expected donut shape (Fig.~1(e)). The THz energy is measured to be about 100 nJ at the entrance to the horn. It should be noted that an azimuthally polarized pulse may also take a donut shape, so a wire grid polarizer is used to confirm the polarization. For a radially polarized pulse, the two lobes after the polarizer line up in the perpendicular direction of the wire (Fig.~1(f)), while those for an azimuthally polarized pulse should line up in the same direction of the wire (see Fig.S4). 

In addition to exciting the accelerating mode, sustained acceleration also requires the THz phase velocity to be equal to the electron velocity, or, in other words, requires the THz frequency to be equal to that of the accelerating mode. In a conventional rf accelerator, this is achieved through precision machining of the rf cavity and utilization of a low-level rf system. Here we vary the crystal temperature to tune the frequency of the THz pulse in order to match that of the accelerating mode. With both the electron beam and THz pulse propagating in the same direction, this THz accelerator works in the traveling-wave mode, and the acceleration process can be visualized through measurement of beam energy change as a function of time delay ($\Delta t$) between the THz pulse and electron beam.

To understand the acceleration process, we schematically depict the relative position of the electron and THz pulses at various time delays in Fig.~2(a). With the THz pulse frequency matching that of the accelerating mode, optimal acceleration should be achieved at $\Delta t=\tau_{THz}=40~$ps when the electron beam catches up with the tail of the THz pulse at the entrance to the DLW. With the beam velocity higher than the group velocity of the THz pulse in the DLW, the electron beam interacts across the full THz envelope while remaining phase matched and achieves maximal energy exchange. Deviating from this time delay causes the beam to miss part of the THz pulse, and the energy exchange is reduced. This is exactly what has been observed in the experiment (Fig.~2(b)), where the envelope of the energy modulation monotonically increases for $\Delta t <40~$ps and then decreases for $\Delta t >40~$ps. Within one wavelength window, the energy change depends on the delay time, which translates to the THz phase seen by the electron beam.

When the THz frequency deviates from that of the accelerating mode, the effective interaction length and energy exchange should be reduced (see Fig.S5). We tuned the THz frequency to 0.51 THz (corresponding dephasing length, i.e. the distance over which the phase slippage increases to $\pi/2$, is about 6.7 mm) to see this effect, and the result is shown in Fig.~2(c). In contrast to the matching case, here at $\Delta t=40~$ps, the energy exchange is close to zero. This is because the electron beam slips from the accelerating phase to the decelerating phase, and the energy gain from interaction with the first half of the THz pulse is canceled by the energy reduction from the second half of the THz pulse, similar to a detuned rf cavity. The simulated electron energy change for the matched and unmatched cases is in excellent agreement with the measurements shown in Fig.~2(d) and 2(e), respectively. Furthermore, no noticeable deflection to the beam has been measured, which indicates that most of the THz energy is converted into a single accelerating mode, demonstrating high efficiency in converting THz energy to accelerating mode and finally to the energy of the electron beam. Here an energy conversion efficiency of about 1.5 keV per nJ$^{1/2}$ has been achieved, much higher than previous results (see Fig.S6). It also allows comparison and confirmation of predictive models and scaling laws and is critical for understanding the physics and engineering issues behind THz acceleration in long structures.

A functioning THz accelerator requires shot-to-shot stability and the staging of the accelerating structures for reaching higher energy. Here a second module 80 cm downstream of the first is used to demonstrate the feasibility, as well as the repeatability of the module. With similar tuning procedures, the energy gain of about 16 keV (for a 110 nJ THz pulse) in the second DLW is achieved. We then vary the time delay of the second THz pulse to change the interaction phase (see Fig.S2), and the corresponding beam energy distribution, as well as transverse beam size (measured at a screen just before the spectrometer magnet), is shown in Fig.~3(a) and 3(b), respectively. From Fig.~3(a), one can see that when the electron bunch interacts with THz pulse at the on-crest phases, its energy is upshifted ($\phi=\pi/2$, acceleration) or downshifted ($\phi=3\pi/2$, deceleration) with minimal broadening. In contrast, at the zero-crossing phases (0 and $\pi$), the beam energy distribution receives maximal broadening with the centroid energy unchanged. Due to the initial beam energy chirp, the broadening at the two zero-crossing phases is slightly different (see Fig.S7). Similar to the zero-phasing method with rf structure \cite{zero-phasing} and laser streaking technique from attosecond metrology \cite{AS}, electron's time information is mapped into its energy distribution when the interaction is around the zero-crossing phases. Accordingly, the electron bunch length is estimated to be about 100 fs RMS and the beam arrival time jitter is determined to be about 50 fs RMS.  

    \begin{figure}[b]
    \includegraphics[width = 0.48\textwidth]{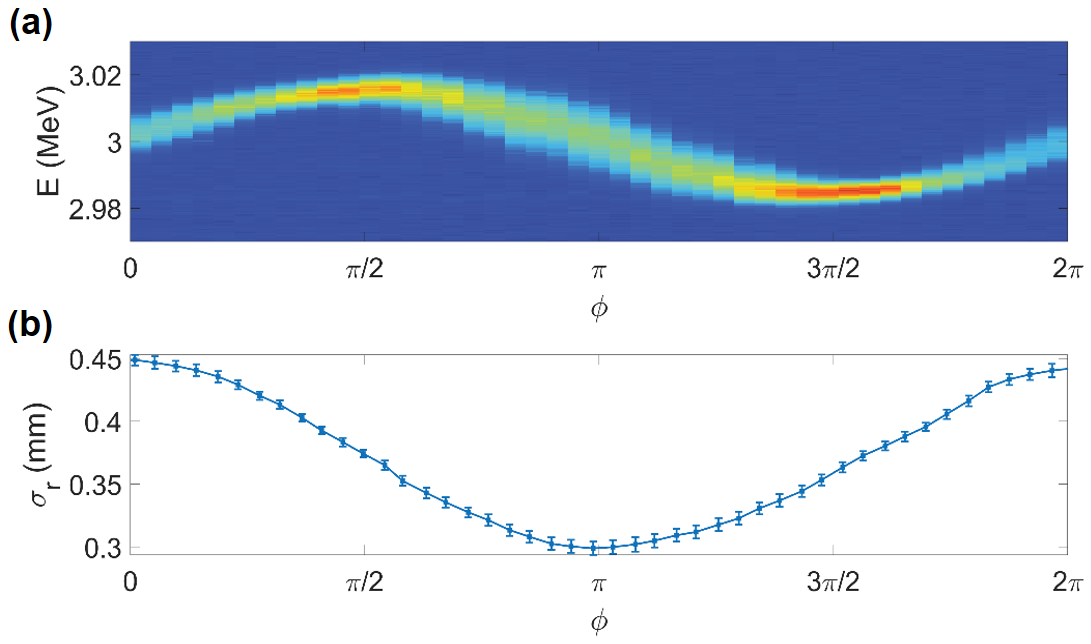}
            \caption{Image of the electron beam on the spectrometer screen (a) and the transverse beam size (b) for various phases of the second THz pulse.   
    \label{Fig.3}}
    \end{figure}  

The beam size in Fig.~3(b) shows the effect of the transverse force in the DLW (see, e.g. \cite{DLW}). Similar to an rf accelerator, the beam experiences a small focusing force at $\phi=\pi$ (bunch head is accelerated more than bunch tail) leading to a reduction in beam size, and a defocusing force is exerted at $\phi=0$ where the bunch head is decelerated more, which is also the phase for bunch compression \cite{RFC1, RFC2}. The transverse force is negligibly small at the accelerating phase (electron beam size is similar to the case with THz off), which is crucial for maximizing the charge coupling efficiency in staging. Due to the absence of transverse force, nearly 100\% charge coupling efficiency is achieved (see Fig.S8).

      \begin{figure}[t]
    \includegraphics[width = 0.48\textwidth]{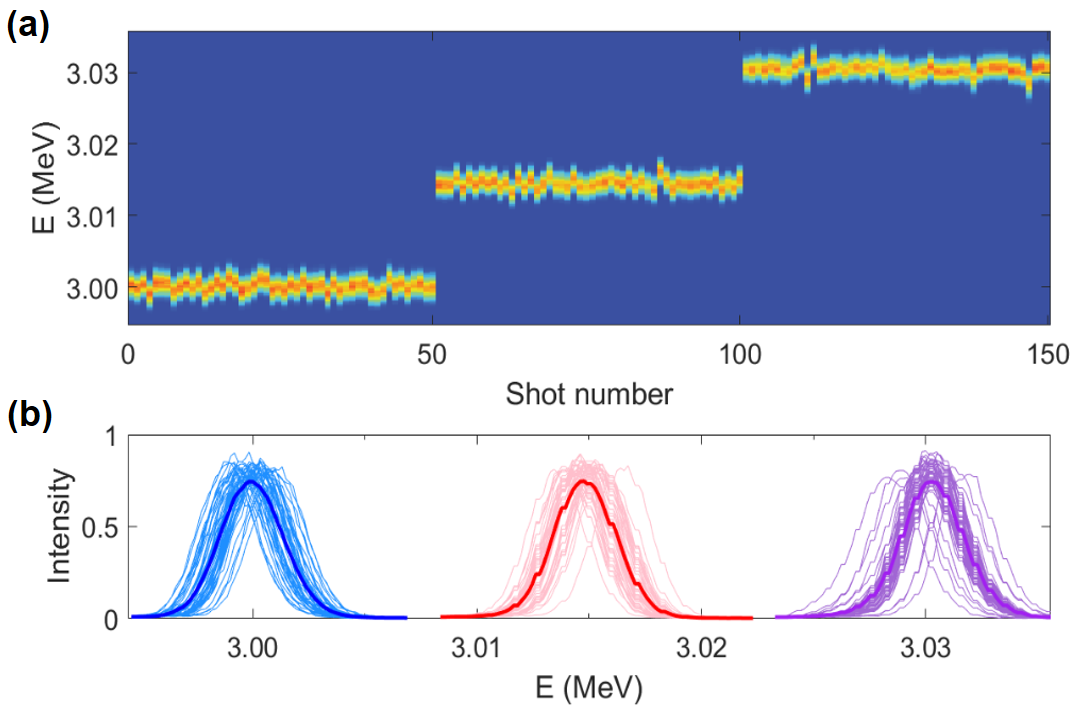}
            \caption{Stable and staged THz-driven interaction. (a) Consecutive measurement of beam energy distribution with both THz pulses off (the first 50 shots), with one-stage acceleration (the middle 50 shots), and with two-stage acceleration (the final 50 shots); (b) Statistical results of the beam energy distribution for the data in (a). 
    \label{Fig.4}}
    \end{figure}  

With the phases of the two THz pulses both set for maximal acceleration, consecutive measurements of the beam energy distribution after one-stage and two-stage acceleration are shown in Fig.~4(a). The corresponding statistical results for beam energy distribution are shown in Fig.~4(b). The beam energy spread (about 1.3 keV) and energy jitter (about 0.6 keV) remain unchanged within the measurement accuracy after staged acceleration, setting a clear route to higher energy with more stages. The shot-to-shot fluctuation of the central beam energy is mainly from fluctuations of rf amplitude in the electron gun, and the fluctuation from the electron beam injection timing jitter is estimated to be about 0.2 keV. 

In summary, we have demonstrated stable and scalable acceleration of a relativistic electron beam in successive DLWs powered by multicycle THz pulses. The energy gain achieved in each module of this mini-accelerator is about 15 keV, limited by the THz pulse energy (about 100 nJ). First, we note that keV electrons have been widely used in ultrafast electron diffraction (UED) for probing nonequilibrium states and structural dynamics \cite{keV1, keV2}. Our technique may immediately find applications in UED, e.g., enabling all-optical UED with matched or even enhanced performance. Second, the DLW may be used as an add-on module for compressing both relativistic and sub-relativistic beams \cite{BaumC, SJTUC, SLACC}, enhancing the performance of UED and 4D electron microscopy \cite{4DEM}. Third, narrowband THz pulse with millijoule-level energy has been demonstrated \cite{HETHz}, so MeV energy gain could be obtained with such a THz source, which opens up new opportunities in fields as portable medical and industrial imaging, and cancer therapy. Furthermore, with the demonstrated stability and scalability, a functioning miniaturized high energy THz accelerator with dozens of stages can be foreseen. 

This work was supported by the National Natural Science Foundation of China (Grants No. 11925505, 12005132, 11504232, and 11721091), the office of Science and Technology, Shanghai Municipal Government (No. 16DZ2260200 and 18JC1410700), and China National Postdoctoral Program for Innovative Talents (No. BX20200220). \\
* These authors contributed equally to this work. \\
$\dag$ dxiang@sjtu.edu.cn \\
$\dag$ jzhang1@sjtu.edu.cn \\

\pagebreak

\end{document}